\documentclass[preprint,epsf,aps,floats]{revtex4}
\usepackage{amsmath,amssymb,bm,epsfig,psfrag}
\renewcommand\d{\partial}
\newcommand\p{{\bm{p}}}

\renewcommand\k{{\bm{k}}}
\newcommand\grad{\bm{\nabla}}
\newcommand\+{\dagger}

\newcommand{\be}{\begin{equation}}
\newcommand{\ee}{\end{equation}}
\newcommand{\ber}{\begin{eqnarray}}
\newcommand{\eer}{\end{eqnarray}}

\newcommand\bra[1]{\langle #1 |}
\newcommand\ket[1]{|{#1}\rangle}

\newcommand\beq{\begin{eqnarray}}
\newcommand\eeq{\end{eqnarray}}

\def\Dsl{\,\raise.15ex \hbox{/}\mkern-12.8mu D}

\begin{document}

\title{Low energy density correlation function of the degenerate unitary Fermi gas
from $\varepsilon$ expansion}
\author{Andrei~Kryjevski}
\affiliation{Nuclear Theory Center, Indiana University, Bloomington, IN~47408 and Department of Physics,~Washington University in St. Louis,~St.~Louis, MO~63130}
\begin{abstract}
Using $\varepsilon$ expansion proposed in~\cite{Nishida:2006br} we 
calculate density  
correlation function of the degenerate Fermi gas
at infinite scattering length to 
next-to-leading order in $\varepsilon$ for excitation energies below quasi particle threshold. 
An expression for the low energy dynamic structure function is derived.
\end{abstract}


\maketitle

\section{Introduction}
Cold dilute gas made of two Fermion species (which will be labeled as ${\uparrow},{\downarrow}$) with two body interactions characterized by 
\ber
n\,a^3\gg 1,~n\,r_0^3\ll1,
\eer
where $n$ is the number density, $a$ is the $s$ wave scattering length and $r_0$ is the effective range of the potential, named "unitary Fermi gas" is one 
of the simplest examples of a strongly interacting Fermi system. Yet, unitary Fermi gas is only a slight idealization of dilute neutron matter 
($a \simeq-18 fm,\,r_0 \simeq 2.6 fm$ in the $^1{\rm S}_0$ chanel) that may exist in the crust of a neutron star. Also, it has been studied extensively in the experiments on 
trapped cold atomic gases with tunable interactions using the Fesh\-bach resonance technique 
\cite{ohara-2002-2002,regal-2004,bartenstein-2004-92,zwierlein-2005-435,zwierlein-2006-311,partridge-2005,partridge-2006-97}. From the 
theoretical point of view, unitary Fermi gas lacks any intrinsic scale and is expected to exhibit universal properties. At the same time theoretical 
description of unitary Fermi gas is difficult due to apparent absence of a small dimensionless parameter which could be used in a perturbative expansion.
 
Inspired by ideas of Nussinov and Nussinov \cite{Nussinov:2004nn}, an analytical technique similar to the $\varepsilon$ expansion in the theory of critical 
phenomena has been recently proposed \cite{Nishida:2006br}. 
The idea is to describe 3 dimensional unitary Fermi gas using perturbation theory in the dimensionality of space, $d$, around $d=4$ and $d=2.$ 
The rational for such an approach is that in $d=2,~4$ unitary Fermi gas simplifies
\begin{itemize}
\item{to the non-interacting Fermi gas in $d=2$,} 
\item{to the Bose-Einstein condensate of non-interacting spin 0 dimers in $d=4$} 
\end{itemize}
\cite{Nussinov:2004nn,Nishida:2006rp}.
Comparison with the results of Monte Carlo simulations suggests that already at the next-to-leading order (NLO) in $\varepsilon$ the expansion around $d=4$ may be a 
useful tool in the description of this system \cite{Nishida:2006br}. While subsequent investigation raised doubts about the convergence of the series in $\varepsilon$ 
\cite{Arnold:2006fr}, one obvious way to check usefulness of the technique is to make predictions for various observable quantities at NLO.
If the trend suggested by the initial NLO results of \cite{Nishida:2006br,Nishida:2006rp,Rupak:2006jj}
were to hold, it would serve as a strong encouragement for further investigation. 

So, as a step in this direction, in this paper we report calculation of the density correlation function of the degenerate unitary Fermi gas in $d=4-\varepsilon$ spatial dimensions to the NLO in $\varepsilon$ for energy below quasiparticle threshold.
Expressions for the low energy dynamic structure function and static structure factor are quoted. The simplest way to make predictions for $d=3$
is to set $\varepsilon=1$ in the NLO expressions (\ref{Skw},\ref{ImSkw},\ref{Sk}). 

\section{The Lagrangian and Definitions}
Due to universality of unitary Fermi gas, 
we choose to work with local four-Fermi interaction with a coupling constant $c_0$ tuned 
to reproduce infinite scattering length.
The Lagrangian density of
the unitary Fermi gas in the presence of the source coupled to density, $\rho(x)$, is 
(here and below $\hbar=1$)
\ber 
\mathcal{L} = \psi^{\dagger}\left(i\d_t + \frac{\grad^2}{2m} + \mu+\rho(x)\right)\psi
 + c_0\,n_\uparrow\,n_\downarrow,
\label{lagrangian}
\eer
where $\psi=(\psi_\uparrow,\psi_\downarrow)^T,$ is spin-1/2 fermion field, and $n_i=\psi^{\+}_i\,\psi_i,$ with
$i={\{}\uparrow,\downarrow{\}}$ are particle density operators, $\mu$ is the particle number chemical potential 
which in this calculation is assumed to be the same for both pairing species. The theory is defined in $d=4-\varepsilon$ spatial dimensions, so that, for example, action is
$${\rm S}=\int{\rm d}t~{\rm d}x^{4-\varepsilon}\mathcal{L}.$$

After Hubbard-Stratonovich transformation, the Lagrangian density (\ref{lagrangian}) takes the 
form
\ber 
\mathcal{L} = \Psi^\+\left(i\d_t + \frac{\sigma_3\grad^2}{2m}\right)\Psi
 + \Psi^\dagger\left[\mu+\rho(x)\right]\sigma_3\,\Psi \nonumber \\ - \frac1{c_0} \phi^*\phi
 + \Psi^\+\sigma_+\Psi\phi + \Psi^\+\sigma_-\Psi\phi^*,
\label{HBlagrangian}
\eer
where $\phi$ is an auxiliary di-fermion field, $\Psi$ is a two-component Nambu-Gor'kov field, $\Psi=(\psi_\uparrow,\psi^\+_\downarrow)^T$,
$\sigma_\pm=\frac12(\sigma_1\pm i\sigma_2),$ with $\sigma_{1,2,3}$ being the Pauli matrices.
In this calculation we will concentrate on the unitary regime, $a=\infty,$ which in the dimensional
regularization corresponds to $1/c_0=0$ \cite{Nishida:2006br}.
It is possible to use $\varepsilon$ expansion to describe system near unitarity \cite{Nishida:2006eu},
and the results of this paper may be extended to the near-unitary regime. This is left for future work. 

The density correlation function is defined by 
\ber 
\mathrm{S}(x)&=&-i\bra{\Omega}{\mathrm{T}\,\psi^{\+}(x)\psi(x)\,\psi^{\+}(0)\psi(0)}\ket{\Omega}
 = i\,\frac{\delta^2}{\delta\,\rho(x)\,\delta\,\rho(0)}\,\mathrm{log\,Z}[\rho(x)]|_{\rho=0},
\label{corrfunctionS}
\eer
where $\ket{\Omega}$ is the ground state, $$\mathrm{Z}[\rho(x)]=\int \mathrm{D}\psi\,\mathrm{D}\psi^{\+}\mathrm{D}\phi\,\mathrm{D}\phi^{*}\,\mathrm{Exp}\,\,{i\int_x\mathcal{L}(\psi,\phi;\rho(x))}$$ is the generating functional and $x=(t,{\bf x}).$

\section{Low Energy Boson Propagator of the Superfluid phase}

Let us review some of the low-energy properties of the superfluid phase of the unitary Fermi gas which will be relevant for the density-density correlator calculation \cite{Nishida:2006rp,Arnold:2006fr}.

For energies below the quasi particle mass, $\Delta\sim\phi_0,$ the only relevant degrees of freedom are the bosonic excitations of the superfluid condensate. The Lagrangian is given by
\ber
\int\,L_{eff}(\phi)
=
-i\,{\rm Tr\,Log}{\left( \begin{array}{ccc}
i\,\partial_t+\frac{\nabla^2}{2\,m}+\mu & \phi_0+g\,{\varphi}(x) \\ \phi_0+g\,{\varphi^*}(x)&
i\,\partial_t-\frac{\nabla^2}{2\,m}-\mu\\\end{array}\right)},
\label{Leff_phi}
\eer
where $\phi_0$ is the uniform medium order parameter which to the NLO in $\varepsilon$ is given by
\ber
\phi_0=2\,\mu_0\left[1+\varepsilon(3\,C-1+{\rm Log}\,2)\right],~C\approx0.14424,~\mu_0=\mu/\varepsilon,
\label{phi0}
\eer 
\cite{Nishida:2006br}, while
the coupling,
\ber
g = \frac{(8\pi^2\varepsilon)^{1/2}}m
  \left(\frac{m\phi_0}{2\pi}\right)^{\epsilon/4},
\label{g}
\eer
was chosen by Nishida and Son \cite{Nishida:2006br} in order to give bosons canonical kinetic term.

Eq. (\ref{Leff_phi}) may be expanded in powers of $\phi$ and its derivatives as well as in powers of 
$\varepsilon$. To NLO in $\varepsilon$ and to the second order in $p_0,\,\epsilon_\p$ we have 
\ber
  L_{eff}=\frac{1}2{\Phi}^{\dagger}
{\mathcal D}^{-1}{\Phi},
\label{Leff_phip0pextend}
\eer 
where ${\Phi}^T=(\phi,\phi^*)$ and  
\ber
{\mathcal D}^{-1}_{11}(p)&=&
{\mathcal D}^{-1}_{22}(-p)={\cal Z}\left(p_0-\frac{\epsilon_\p}2\right)+
2\,\mu+
\frac{\varepsilon}{\phi_0}\,\left(\frac{5\,p_0^2}{24}-\frac{p_0\,\epsilon_\p}{6}+
\frac{7\,\epsilon_\p^2}{120}\right)\nonumber \\
&-&
\frac{3\,\varepsilon\,\phi_0}2+
{\cal O}\left(\varepsilon^2,\,\varepsilon\frac{p0^3}{\phi_0^2},\,\varepsilon\frac{\epsilon_\p^3}{\phi_0^2}\right),\nonumber \\
{\mathcal D}^{-1}_{12}(p)&=&{\mathcal D}^{-1}_{12}(p)=-\frac{\varepsilon\,\phi_0}2+\frac{\varepsilon\,\epsilon_\p}{8}-
\frac{\varepsilon\,\epsilon_\p^2}{40\,\phi_0}-
\frac{\varepsilon\,p_0^2}{24\,\phi_0}+
{\cal O}\left(\varepsilon^2,\,\varepsilon\frac{p0^3}{\phi_0^2},\,\varepsilon\frac{\epsilon_\p^3}{\phi_0^2}\right),
\label{d}
\eer
with $p=(p_0,\p),~\epsilon_\p=\p^2/2m,~{\cal Z}=1+\varepsilon({\rm log}~2-\gamma)/2,$ where $\gamma\approx0.57722$, is the Euler-Mascheroni constant.
To NLO in $\varepsilon$ the low-energy effective Lagrangian only contains quadratic terms in $\phi.$ The non-Gaussian terms are suppressed by higher powers of $\varepsilon.$ The situation is analogous to the chiral perturbation theory of the Color-Flavor-Locked phase of asymptotically dense quark matter where meson interaction terms are exponentially suppressed~\cite{Zarembo:2000pj}. 
A generic term in the NLO Lagrangian is of the form 
${\Phi}^{\dagger}c_{rs}\,\varepsilon~p_0^r\epsilon_\p^s/\Delta^{r+s-1}{\Phi},~r+s\geq1,~c_{rs}$ is ${\cal O}(1);$ $p_0,\epsilon_\p\leq\Delta$ are treated as ${\cal O}(1)$ in $\varepsilon$ expansion.

In the $\varepsilon$ expansion approach observables are calculated in the perturbation theory where fermions interact with bosonic excitations of the condensate with the coupling $g\propto \sqrt{\varepsilon}$ of (\ref{g}). Detailed descriptions of Feynman rules at both zero and finite temperature are presented in \cite{Nishida:2006rp,Nishida:2006eu}.

Superficially, to calculate density correlator to the NLO in $\varepsilon$ it should be sufficient to use the leading order (LO) boson propagator which is proportional to $(p_0-\epsilon_\p/2)^{-1}.$ 
However, it does not have the superfluid mode pole at $p_0=c_s\,|\p|$ as $|\p|\rightarrow 0$, where $c_s$ is the speed of sound,
which we know should be present in the spectrum. It has been shown that to capture the right form of the superfluid mode excitation energy 
it is sufficient to use boson propagator with momentum independent ${\cal O}(\epsilon)$ boson self-energy corrections resummed~\cite{Nishida:2006rp,Arnold:2006fr}. Then
the relevant NLO boson propagator is
\ber
{\mathcal D(p)}_{11}=\frac{p_0+\frac{\epsilon_\p}{2}}{p_0^2-\epsilon_\p(\frac{\epsilon_\p}{4}+\mu)+i\delta},~\delta= 0^{+}.
\label{D11}
\eer

We will need sound velocity to NLO. From the NLO boson propagator (\ref{Leff_phip0pextend}) and (\ref{d}) using 
(\ref{phi0})
one gets
\ber
c_s^2=\frac{\mu}{2\,m}\left[1+\frac\varepsilon4\right].
\label{phononNLO}
\eer

\section{The Density Correlation Function}
\begin{figure}[t]
\centering{
\begin{psfrags}
\psfrag{c}{${\cal{O}}(1/\varepsilon)$}
\psfrag{a}{${\cal{O}}(1)$}
\psfrag{g}{${\rm g}$}
\psfrag{mu}{$\mu$}
\psfrag{rho}{$\rho(x)$}
\psfrag{rho0}{$\rho(0)$}
\epsfig{figure=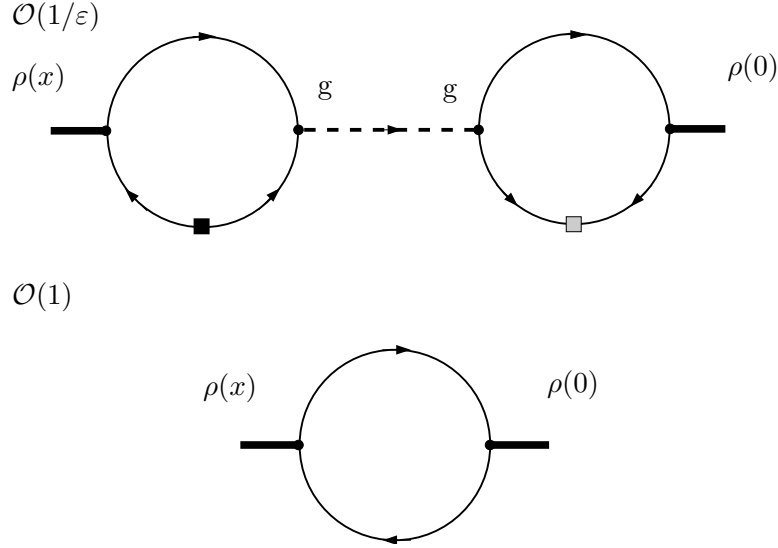, width=.6\textwidth}
\end{psfrags}
}
\caption{Feynman diagrams relevant for the calculation of the density correlation function to NLO in $\varepsilon$.
Thin solid lines depict propagating fermions in the superfluid background (lines with squares correspond to the anomalous propagators), dashed line is the dressed boson propagator, thick solid line is the source coupled to the particle density. The ${\cal{O}}(1/\varepsilon)$ diagram contributes to the imaginary part of the density-density correlator below quasiparticle threshold, $\omega < 2\,\Delta,$ where $\Delta\sim\phi_0$ is the quasiparticle mass. 
}
\label{fig:graphs}
\end{figure}

To NLO in $\varepsilon$ the density-density correlation function will consist of the contributions from Fig. (\ref{fig:graphs}): the one boson exchange diagram which produces 
${\cal O}(1/\varepsilon)$ contribution, and the ${\cal O}(1)$ one fermion loop diagram.

The dashed line in Fig. (\ref{fig:graphs}) is the dressed boson propagator discussed in the previous section (Eq. (\ref{D11})). Thin solid line represents fermion propagator 
\ber
G(p_0,\p) = \frac1{p_0^2-E_\p^2+i\delta}\left(
   \begin{array}{cc}
      p_0 + \epsilon_\p - \mu& -\phi_0 \\
      -\phi_0 & p_0-\epsilon_\p + \mu
 \end{array}
  \right),~\delta= 0^{+}
\label{S}
\eer
where 
$E_\p=((\epsilon_\p-\mu)^2+\phi_0^2)^{1/2}$. The thick solid line is the source $\rho(x)$ coupled to the particle density. The coupling $g$ is given in Eq. (\ref{g}). 
For energies below twice the quasiparticle gap the only contribution to the imaginary part of density-density correlator will come 
from the one boson exchange diagram.

The one boson exchange diagram and the one loop graph contributions are given by
\ber
S_{NGB}(k)&=&
-\int_{p,\,q}\,g^2\,
{\rm{tr}}\left[G(p)\,\sigma_3\,G(p+k)\,\sigma_{-}\right]\,{\mathcal D(k)}_{11}\,
{\rm{tr}}\left[G(q)\,\sigma_3\,G(q-k)\,\sigma_{+}\right]+\nonumber \\&+&
(k\leftrightarrow-k),
\label{Dkw}
\eer
and
\ber
\mathrm{S}_{1 \,loop}(k)=-i\,\int_p\,{\rm{tr}}
\left[G(p)\,\sigma_3\,G(p-k)\,\sigma_3\right],
\label{Sden1loop}
\eer
respectively, and $\int_k\equiv\int\,d^{d+1}k/(2\,\pi)^{d+1}.$
The fermion loop integrals are calculated in $d=4-\varepsilon$ dimensions expanding to the second order in 
$k_0,\,\epsilon_k,$ and $\sigma_{\pm}, \sigma_3$ were defined below Eq. (\ref{HBlagrangian}). It was argued in \cite{Nishida:2006br} that $\mu$ was to be counted as ${\cal O}(\varepsilon).$
To the NLO $\mu\sim\varepsilon$ is omitted everywhere in (\ref{Dkw}) except in the denominator of ${\mathcal D(k)}_{11}.$ 

The result normalized to the NLO particle density, $n,$ \cite{Nishida:2006br}
is
\ber
\frac1nS(p_0,\epsilon_\p)&=&
\frac{2\,\epsilon_\p\left(1-\frac{\varepsilon}4\right)-\frac{{\varepsilon}}{6\,\mu_0}
\left(6\,p_0^2+\epsilon_\p^2)\right)+{\cal O}\left(\varepsilon^2,\,\varepsilon{\epsilon_\p^3}/{\mu_0^2}\right)}
{p_0^2-{\epsilon_\p}(\mu+\frac{\epsilon_\p}{4})+i\delta+{\cal O}\left(\varepsilon{\epsilon_\p}{\mu}\right)} 
-\nonumber \\&-&
\frac{\varepsilon}{4\,\mu_0}\left(1-\frac{\epsilon_\p}{6\,\mu_0}+\frac{p_0^2}{24\,\mu_0^2}+
\frac{\epsilon_\p^2}{120\,\mu_0^2}\right)+{\cal O}\left(\varepsilon^2,\,\varepsilon\frac{p0^3}{\mu_0^4},\,\varepsilon\frac{\epsilon_\p^3}{\mu_0^4}\right),\nonumber \\&& 0 \leq p_0 \leq 2\Delta,\,\delta= 0^{+},
\label{Skw}
\eer
where
\ber
n=\left(\frac{m\,\mu_0}{2\,\pi}\right)^{d/2}\,\frac{4}{\varepsilon}\,
\left[1+\varepsilon\left(6\,C-\frac{\gamma_E}{2}+2\,{\rm Log}\,2-\frac{7}{4}\right)\right],~
\gamma_E\approx0.57722, ~C\approx0.14424,
\label{n}
\eer
$\mu_0=\mu/\varepsilon\sim\phi_0,$ and $\Delta/\mu_0\simeq 2(1-0.345\varepsilon)$ is the NLO quasiparticle mass \cite{Nishida:2006br}, and the NLO expression for the uniform medium superfluid order parameter (\ref{phi0})
has been used.
The boson exchange contribution to (\ref{Skw}) is not fully expanded in the powers of $p_0,\,\epsilon_\p$ to preserve the pole structure. 
The dynamic structure factor, $\sigma(\omega,\epsilon_\p)$, is, then
\ber
\sigma(p_0,\epsilon_\p)\equiv-{\rm Im}\,S(p_0,\epsilon_\p)&=&
\pi\,n\,\left[\frac{2\,\epsilon_\p\left(1-\frac{\varepsilon}4\right)-\frac{{\varepsilon}}{6\,\mu_0}
\left(6\,p_0^2+\epsilon_\p^2)\right)}{2\,p_0}\right]\times\nonumber \\&\times&
\delta\left(p_0-\sqrt{\epsilon_\p\left(\mu+\frac{\epsilon_\p}{4}\right)}\right),
\label{ImSkw}
\eer
where $n$ is the NLO equilibrium density of Eq. (\ref{n}). And the inelastic form factor (per unit volume), 
${\cal{S}}_{inel}(\epsilon_\p)$, in this approximation is given by
\ber
{\cal{S}}_{inel}(\epsilon_\p)\equiv-\frac{1}{\pi\,n}\int_{0}^{\infty}{\rm d}\omega\,{{\rm Im}\,S(\omega,\epsilon_\p)}&=&
2\,\sqrt{\frac{\epsilon_\p}{\epsilon_\p+4\,\mu}}\left[1-\frac\varepsilon4-\frac{5\,\varepsilon\,\epsilon_\p}{24\mu_0}\right].
\label{Sk}
\eer

\section{Sum Rule Checks}

Let us check the correlation function (\ref{Skw}) against several 
identities for a density correlator. 
Using $S(k)$ of (\ref{Skw}) we find that
\begin{enumerate}
\item{the dispersion relation
\ber
S(\omega=0,\epsilon_k)=\frac{2}{\pi}\int_{0}^{\infty}{\rm d}\omega \frac{{\rm Im}\,S(\omega,\epsilon_k)}{\omega}
\label{dispersion_relation}
\eer
is satisfied to the NLO in $\varepsilon$ and to ${\cal O}(\epsilon_\k^2);$}
\item{the compressibility sum rule
\ber
S(\omega=0,\epsilon_\k=0)=-\frac{n}{m\,c_s^2}
\label{compress}
\eer
where $n$ is the equilibrium density given by (\ref{n}) and $c_s$ is the speed of sound given by (\ref{phononNLO}) is satisfied to the NLO in $\varepsilon.$ Note that the l.h.s. of (\ref{compress}) only contains the LO expression for $c_s.$ One also observes that the $1-\varepsilon/4$ coefficient at $\epsilon_\p$ in the numerator of the fraction in (\ref{Skw}) is necessary to satisfy (\ref{compress}) to NLO;}
\item{the energy weighted sum rule which for a unit volume of the medium
is
\ber
-\frac{1}{\pi\,n}\int_{0}^{\infty}{\rm d}\omega\,\omega\,{{\rm Im}\,S(\omega,\epsilon_\p)}=\epsilon_\p
\label{energy_weighted}
\eer
for any $\epsilon_\p.$
From $S(\omega,\epsilon_\p)$ of (\ref{Skw}) we get 
\ber
-\frac{1}{\pi\,n}\int_{0}^{\infty}{\rm d}\omega\,\omega\,{{\rm Im}\,S(\omega,\epsilon_\p)}=
\epsilon_\p\left[\left(1-\frac\varepsilon4\right)-\frac{5\,\varepsilon\,\epsilon_\p}{24\mu_0}\right].
\label{energy_weighted_eps}
\eer
Let us emphasize that sum rule (\ref{energy_weighted}) is sensitive to the dynamics of quasi particles at 
$\omega>2\,\Delta$ which we have not included in the calculation.
We see that the sum rule is satisfied at LO in $\varepsilon.$ At NLO the 
$\epsilon_\p$ coefficient is reduced to $1-\varepsilon/4$ instead of 1 prescribed by (\ref{energy_weighted}) which is due to the omission of the quasi particle contribution above the $2\,\Delta$ threshold. Also higher order terms 
suppressed by powers of $\epsilon_\p/\mu_0\sim\epsilon_\p/\Delta$ appear which is natural within the framework of the low energy effective theory. So, we see that including only the condensate excitation contribution to the NLO
${{\rm Im}\,S(p)}$ and omitting the rest of the (heavy) intermediate states we do not violate the energy weighted sum rule (\ref{energy_weighted}) too badly.}
\end{enumerate}
These checks indicate that 
the density-density correlator given by 
Eq. (\ref{Skw}) reproduces response properties of the medium at low energy reasonably well.

\section{Conclusions}
We have computed density correlation function of unitary Fermi gas, $S(\omega,\epsilon_\p),$ to NLO in the $\varepsilon$ expansion for energies below quasi particle threshold 
$\omega \leq 2\Delta$ and quoted expression for low energy dynamic structure function and static structure factor. The simplest extrapolation to $d=3$
may be made by simply setting $\varepsilon=1$ in the NLO expressions (\ref{Skw},\ref{ImSkw},\ref{Sk}). As is well known (see, ${\it e.}{\it g.},$ \cite{Negele:elastic}), dynamic structure function is proportional 
to the inelastic scattering cross-section of any external probe coupled to density, and may be measured experimentally.

Based on the calculation one predicts a significant change in the dynamic structure function at the quasiparticle threshold, $\omega = 2\Delta,$ which may help experimental 
measurement of the quasi particle gap.

Future work will include calculation of the dynamic structure function for energies above quasi particle threshold and of the spin response function relevant for the description of neutrino coupling to the dilute neutron matter (${\it e.}{\it g.},$  \cite{Reddy:1998hb,Horowitz:2006pj}).
\section{Acknowledgment}
The author is grateful to C.J.~Horowitz for suggesting the problem and numerous helpful discussions and to Sanjay Reddy for pointing out significance of the boson exchange diagram of Fig.~(\ref{fig:graphs}). Also the author would like to thank 
Yusuke Nishida and Dam Son for useful conversations. This
work is supported in part by the US Department of Energy grants DE-FG02-87ER40365 and DE-FG02-05ER41375 (OJI) and by the National Science Foundation grant PHY-0555232.
\bibliography{cfl}

\end{document}